\documentstyle[preprint,aps,epsfig]{revtex}

\newcommand\be{\begin{equation}}
\newcommand\ee{\end{equation}}

\newcommand {\cS   } {{\mathcal{S}}         }

\begin{document}

\title{Protein Design is a Key Factor for Subunit-subunit Association}
\author{Cecilia Clementi$^{1,\dag}$ \footnote{Corresponding author: e-mail
address: cecilia@curio.ucsd.edu}, Paolo Carloni $^{1,2}$, Amos Maritan$^{1,3}$
 }
\address{$^1$ International School for Advanced Studies (SISSA)}
\address{and Istituto Nazionale di Fisica della Materia,}
\address{Via Beirut 2-4, 34014 Trieste, Italy}
\address{}
\address{$^2$International Centre for Genetic Engineering and\\
Biotechnology, I-34012 Trieste, Italy}
\address{$^3$ Abdus Salam International Center for Theoretical Physics,}
\address{Strada Costiera 11, 34014 Trieste, Italy}
\address{}
\address{$^{\dag}$Present Address: \\Department of Physics, University of California at San Diego, \\ La Jolla, California 92093-0319, USA}
\date{\today }
\maketitle

\vspace{1cm}

\newpage

{\bf Abstract}

Fundamental questions about  role of the quaternary structures
are addressed using a statistical mechanics off-lattice model of a dimer protein.
The model, in spite of its simplicity, captures key features of the
monomer-monomer interactions revealed by atomic force experiments. Force
curves during association and dissociation are characterized by sudden jumps
followed by smooth behavior and form hysteresis loops. Furthermore, the
process is reversible in a finite range of temperature stabilizing the
dimer. It is shown that in the interface between the two monomeric subunits
the design procedure naturally favors those amino acids whose mutual
interaction is stronger. Furthermore it is shown 
that the width of the hysteresis loop
increases as the design procedure improves, i.e. stabilizes more the dimer.

\newpage

\section{Introduction}

\label{data.multi.sec}

Molecular recognition is a process by which two biological molecules
interact to form a specific complex. Structural domains of proteins
recognize ligands, nucleic acid and other proteins in nearly all fundamental
biological processes. The recognition comprises a large spectrum of specific
non bonded interactions, such as van der Waals interactions, hydrogen
bonding and salt bridges, which overcame the loss of conformational entropy
upon association\cite{book_proteins}. These interactions are ubiquitous, yet
they are responsible for the exquisite specificity of the aggregation.
Understanding the aggregation process is important not only for our
comprehension of the formation of the molecular aggregates but also to gain
insights on how the interactions cancel each other in the many other
possible supramolecular modes of association. Investigating the nature of
association can also shed light on the protein folding processes, as the
aggregation process can be described as the transfer of surface from water
to the protein interior. Useful information could also be obtained on the
complex relationship between cooperativity and quaternary structure in
proteins such as myoglobin. Finally, such type of studies can have important
application in pharmacology and medicine. A typical example in this respect
is insulin, a protein whose efficiency for the treatment of
insulin-dependent diabetes could be boosted by a better understanding of the
association/dissociation mechanism\cite{whit}. A deep understanding of the
aggregation mechanism is of current interest also in anti-AIDS research.
Subunit-subunit association inhibitors of the dimeric enzyme HIV-1 protease
are currently putative agents against HIV infection\cite{hiv_rev}. These
drugs are ligands with high affinity at the interface region.

While several biochemical and biophysical experiments have been directed
toward characterizing thermodynamic \cite{book_proteins,td} and kinetic\cite
{kin} aspects of protein-protein interactions, a small number of experimental%
\cite{banci} and theoretical\cite{vang} studies have addressed the role of
subunit association at the atomic level.

In this respect, an elegant experiment is represented by the determination
of the monomeric structure of a dimeric enzyme, Cu, Zn- superoxide dismutase
obtained by mutating key residues at the subunit-subunit interface\cite
{banci}. The three dimensional structure of the single subunit exhibits
small differences with the native enzyme, yet the conformation is less
favorable of the substrate to the reaction site, indicating an essential
role of the quaternary structure.

Subunit-subunit interactions have been recently measured by direct
force measurements. Yip et al.\cite{Bio98} have revealed the
complexity of the insulin dimer dissociation and have shown the
energetics associated to the disruption of discrete molecular bonds at
the monomer-monomer interface.  While these studies are capable to
quantify the forces governing protein aggregation, the fundamental
question on how the ${\it design}$ quality of proteins (namely its
topology and its three dimensional structure) affects domains of
different subunits has never been addressed. Here, we use a simple
statistical mechanical model to investigate the relationship between
the design of a dimeric protein and the interactions present the
subunit-subunit interface. Comparison is made with experiments such as
those of Yip et al.\cite{Bio98} by calculating the forces necessary to
pull away and push back the two subunits. We find that our model not
only is able to capture key aspects of these experiments, but also
provides novel information on the intricate relationship between the
design of a dimeric protein and the interactions present the
subunit-subunit interface.

%-------------------------%

\section{Computational Section}

{\it Energetics}\cite{units}.We use a very simple protein model in which only
the $C_\alpha $ atoms are considered \cite{cmb98}, and a very simple form of
the monomer-monomer interaction energy: 
\begin{equation}
V=V^A+V^B+V^{AB},  \label{pot_tot}
\end{equation}
where $V^A$ ($V^B$) is the potential energy of $N_A$ ($N_B$) interacting
beads constituting chain $A$ ($B$): 
\begin{equation}
V^A=\sum_{i<j\;\;i=1,N_A}\left\{ \delta _{i,j+1}f(r_{i,j})+\eta
(a_i,a_j)\left[ \left( \frac \sigma {r_{ij}^A}\right) ^{12}-\left( \frac 
\sigma {r_{ij}^A}\right) ^6\right] \right\} ,  \label{potA}
\end{equation}
and $V^{AB}$ is the potential energy given by the interaction of beads of
chain $A$ with beads of chain $B$: 
\begin{equation}
V^{AB}=\sum_{i=1,\cdots ,N_A\;\;j=1,\cdots ,N_B}\left\{ \eta (a_i,a_j)\left[
\left( \frac \sigma {r_{ij}^{AB}}\right) ^{12}-\left( \frac \sigma {%
r_{ij}^{AB}}\right) ^6\right] \right\} .  \label{potAB}
\end{equation}
In eq.~(\ref{potA}) $r_{ij}^A=|{\bf r}_i^A-{\bf r}_j^A|$ represents the
distance between beads $i$ and $j$ of chain A, whereas in eq.~(\ref{potAB}) $%
r_{ij}^{AB}=|{\bf r}_i^A-{\bf r}_j^B|$ is the distance between bead $i$ of
chain A and bead $j$ of chain B. The parameters $\eta (a_i,a_j)$ and $\sigma 
$ entering these equations determine, respectively, the energy scale and the
interaction range between monomer kinds $a_i$ and $a_j$. Twenty different
beads are used to represent the natural amino acids present in proteins.
Interaction parameters between each amino acid pair $i,j=1,\dots ,20$ are
chosen to be all attractive (i.e. $\eta (a_i,a_j)>0\;\forall (i,j)$) and
proportional to the values determined by Miyazawa and Jernigan (MJ) \cite{MJ}%
. The function $f(x)$ in eq.~(\ref{potA}) represents the energy of the
virtual $C_\alpha $--$C_\alpha $ peptide bond and it is equal to: 
\begin{equation}
f(x)=a(x-d_0)^2+b(x-d_0)^4,  \label{C_C_bond.c6}
\end{equation}
with $a$ and $b$ taken to be $1$ and $100$ respectively, and $d_0$ set equal
to $3.8$ \AA . The effect of $f(x)$ is to act as a ``soft clamp'' to keep
subsequent residues at nearly the typical distance observed in real proteins.

%-------------------------%

{\it Protein model.}{\bf \ }We construct a model of a dimeric protein with a
given symmetry $\cS$ by selecting a compact, low--energy configuration. To
build the $3-D$ structure we follow the procedure of ref.~\cite{cmb98}. An
homopolymer is collapsed through molecular dynamics with a potential in Eq.(%
\ref{pot_tot}) where all the $\eta $'s are taken equal to their maximum
(i.e. the most attractive) value $\eta _{max}=10$, and $\sigma $ is chosen
to be equal to $\sigma _0=6.5\AA $. In practice we consider the motion of
chain $A$ only under the potential: 
\begin{equation}
V=V^A+\frac 12V^{A\;\cS (A)},
\end{equation}
where $\cS (A)$ is the chain configuration obtained from of the application
of the symmetry transformation $\cS$ to the chain $A$. The potential $V^{A\;%
\cS(A)}$ depends only on the coordinates of chain $A$.
In the case of a $C2$ symmetry, if $%
\;(x_i,y_i,z_i)\;\;i=1,\cdots ,N_A$ are the coordinates of beads of chain $A$%
, then $\;(-x_i,-y_i,z_i)\;\;i=1,\cdots ,N_A$ are the coordinates of those
in chain $\cS (A)$ and the expression of $V^{A\cS (A)}$ is: 
\begin{equation}
V^{A\;\cS(A)}=\sum_{i,j=1,\cdots ,N_A}\{\eta ((\frac{\sigma _0}{\tilde{r}%
_{ij}})^{12}-(\frac{\sigma _0}{\tilde{r}_{ij}})^6)\}
\end{equation}
where $\tilde{r}_{ij}=\sqrt{(x_i+x_j)^2+(y_i+y_j)^2+(z_i-z_j)^2}$.

The target conformation of the dimer (''native state'') is obtained by collapsing
a randomly generated swollen configuration of a chain made up of $50$ monomers
by using Molecular Dynamics (MD) simulations combined with a slow
cooling procedure. The procedure is repeated several times from different
random initial conditions (for details see \cite{cmb98}).

{\it \ }The procedure of ref.~\cite{cmb98} is used to assign a suitable
sequence to the selected structure. We used twenty kinds of amino acids $%
a_i\;\;i=1,\cdots ,20$, with the values of the matrix elements $\eta
(a_i,a_j)\;\;i,j=1,\cdots ,20$ as given in ref. \cite{MJ}. The compositions
of both the sequences (of chain $A$ and $B$) --shown in Table \ref{table2}
are chosen as the ones occurring in real proteins \cite{klapper}.

Following ref. \cite{cmb98} small variations in the Lennard--Jones length
parameter are allowed. Here 6 possible values ($\sigma _0=6.5$, $\sigma _1=6$%
, $\sigma _2=6.25$, $\sigma _3=7.0$, $\sigma _4=7.5$ and $\sigma _5=8.0$ \AA
) are permitted, both for potential $V^A$ ($V^B$) and $V^{AB}$. The
three-dimensional structure and the intra-- and inter-- chain contacts as
obtained from our design procedure are shown in Fig. 1. Forces as a function
of the distance {\it r } of the centers of mass of the two subunits are
calculated {\it i)} by positioning two monomers at the distance {\it r} with
the conformation of the native protein (''static force''); {\it ii)} by
performing constrained Molecular Dynamics (CMD) simulations \cite{smit}\cite
{rattle}\cite{shake}, in order to equilibrate the system 
with {\it r } as constraint coordinate.

\section{Results}

{\bf Energy landscape and folding}.{\bf \ }Almost all dimeric globular
proteins exhibit symmetric or pseudo symmetric aggregation. Due to the
amino acid chirality, only symmetry point groups not containing the
inversion are possible. By far the most common point group symmetry is
the $C2$. Examples range from HIV-1 protease to Cu,Zn-superoxide dismutase,
immuglobulin and glutathione reductase and many others
\cite{book_proteins}.We therefore impose a $C2$ symmetry to the native
state of the model dimeric protein. The target structure (''native
state'') is chosen as the lowest--energy conformer for the two
subunits equipped with the designed sequences, $A$ and $B$ (see
Computational Section).

To explore the energy landscape of the dimeric protein, we first collected
several configurations of the subunits during MD simulations.

Fig. 2 shows the energy of each configuration of the two chains $A$ and $B$
plotted against the distance,{\it d},from the dimer native state. The
distance, among configurations is measured by 
using the Kabsch expression \cite{kab}.

The two subunits are able to find the lowest minimum in a reasonable time
(e.g. in a time observable by a MD simulation -- typical simulation runs are
about 10.000 $\tau $ ) only if starting from a set of initial conditions.
Obviously, if all beads of chain $A$ are initially located too far --with
respect to the Lennard--Jones interacting distances $\sigma $-- from any
bead of chain $B$, the two chains would evolve independently, and the dimer
structure could not be reached.

{\bf Subunit-subunit forces. }These interactions are experimentally measured
by attaching part of the subunits to opposing surfaces and measuring the
resulting forces at several subunit-subunit distances \cite{Bio98}. To mimic
the anchoraging of the subunits, we carry out MD simulations in which the
distance {\it r} between the centers of mass of the subunits is constrained.

Fig. 3 plots the constrained force as a function of{\it \ r }at a
fixed temperature (=0.06 $\varepsilon $)\cite{units}.  Both the force
required to dissociate the dimer and that to reform it are
reported. Several interesting features emerge from this graph. First,
both association and dissociation forces experience sudden jumps
followed by a smooth behavior of the
force\cite{Bio98,allm1,allm2,allm3,allm4,allm5}.\ Second, by pulling
and pushing back the subunits, an hysteresis-like circle is formed.
Both behaviors are observed also in experiments of protein
dimers\cite {Bio98,allm1,allm2,allm3,allm4,allm5}. Consistently, a
careful analysis of the contacts between the aminoacids of the
different subunits indicates that the number of contacts destroyed
during dissociation at a given intermonomer distance {\it r} is {\it
not} equal to those formed during the association.

Finally, we notice that the ''static'' force, namely the force calculated
leaving the two subunits in their original native conformation,
is remarkably different, stressing
the role of protein relaxation in subunit-subunit interactions.

{\bf \ Protein Stability. }To analyze the interactions at the subunit
interface and compare them with those stabilizing the interior of the
protein we calculate the distribution of the contacts \cite{Cec_Michele}
as a function of their
energies at several subunit-subunit distances (Fig. 4). In the native state,
the design procedure naturally  favors stronger forces across the
monomer-monomer interfacial contacts relative to the intra-chain ones. This
is also in agreement with the suggestions coming from the experiments of Yip
et al. \cite{Bio98}. As $r$ increases the weakest inter-monomer bonds are
broken and the strength distribution becomes more and more peaked around the
strongest interaction. In contrast, the intra-chain bonds distribution
remains almost unaltered as a results of the rupture of some contacts and
formation of new ones.

{\bf Protein Deformation. }Our calculations allow also for characterizing
the deformation of the monomers as the dimer is pulled. The deformation is
measured as the distance (by using the Kabsch expression \cite{kab}) of each
monomer from the configuration it had in the native state, averaged over the
two monomers (the values for the two monomers turn to be quite similar).

Fig. 5(a) ~ shows the results of three simulations at temperature 
$T=0.05\epsilon $ when the dimer is slowly pulled and then slowly released.
It is visible that the final states are always the same but the pathways to
reach them can be quite different. This is fully in agreement with the
energy landscape theory and the protein folding funnel concept 
\cite{Leopold92,Betancourt95,Bryngelson95,Onuchic95,Onuchic96,Onuchic97,Nymeyer98}
.This behavior persists also at higher temperatures e.g. $T=0.10\epsilon $.
However, at more higher temperature after the dimer is released the native state
in not more recovered and another locally stable configuration is reached
($T= 1.15 \epsilon$ is shown in Fig. 5(b)). 
A similar situation occurs also at very low temperatures (Fig.
5(c)). After the dimer is pulled the system is unable to reach again its
native state when it is released. This indicate that the process of pulling
and releasing is `` reversible'' only in a certain range of temperature. The
temperature T has not to be too high (i.e. T has to be lower than a folding
temperature $T_f$) otherwise the system escape from its local minimum and
not too low (i.e. T has to be higher than a glassy temperature $T_g$)
otherwise thermal fluctuations are not enough to allow the system to
overcome the small scale roughness of the energy landscape
(see for instance \cite{Onuchic95,Onuchic96,Onuchic97}).

{\bf Monomers' refolding. }In our ideal dimer, the primary structures for 
{\it A }and {\it B }are very similar but not equal. This is done to study
the refolding process independently on the two monomeric subunits. Starting
from swollen configurations well separated, the two monomers refold as two
independent chains. We collected several configurations during MD
simulations of these independent monomers. The minimum energy conformations
of the couple of non--interacting monomers turns out to be composed by very
similar $A$ and $B$ monomeric structures ($0.7$ \AA\ of RMS difference) and
not too different from the monomer structures constituting the dimer native
state (about $2-3$ \AA\ each). The lowest energy conformations of the two
non--interacting monomers are those reached at the final stage of the
pulling process.

The similarity between the stable conformations of chains $A$ and $B$ folded
separately could be predicted since their sequences are similar, although
not identical.

The similarity between the independently folded conformations and those in
the native protein agrees with the experimental results \cite{banci}. A conformational
drift is experienced, as described before. Indeed, as Figure~6 illustrates,
the refolded minimum--energy structure differs from the native one only for
a rearrangement of the interfacial region.

{\bf Design quality. } We have also addressed the question of how the width of 
the hysteresis cycle observed in fig. 3 depends
on the design quality of sequences A and B. To further simplify matter
we have restricted the number of amino acids classes to four, without 
loss of generality. The same structure of the dimer and the same design 
procedure \cite{cmb98} as above have been employed. However the design 
procedure was applied twice, one more and the other less accurate,
providing two different pairs of sequences with the same native state 
and two different degrees of stability.
The association and dissociation forces for these two new cases are 
shown in Fig. 7 and clearly indicate the strong correlation between 
the design quality and the width of the hysteresis loop.

\section{Discussion and Conclusion}

Our model of a dimeric protein, in spite of its extreme simplicity, is able
to capture several specific features of the monomer/monomer interactions.
Indeed, it reproduces the experimental ''jump'' behavior of the
subunit-subunit force, which has been suggested to be a consequence of the
disruption of a hierarchy of different kinds of intermolecular interactions
(hydrogen bonds, electrostatic and dispersion forces and salt bridges) at
the monomer--monomer interface. Furthermore, it provides hysteresis-like
circles as observed in direct force experiments on dimeric proteins.
Consistently, we find that the number of contacts between aminoacids at the
interface is different in the association and  in the formation processes.
Finally, our calculations indicate that the native state of a dimeric
protein is recovered if the two subunits are pushed back, again in agreement
with experimental evidence\cite{Bio98}. The agreement with experiment is
rather striking considering that it is achieved by designing the dimeric
protein with a extremely simple force field (a Lennard--Jones potential)
mimicking the complex interactions among amino acids. This strongly suggests
that the characteristic feature of inter--monomer force curves can be simply
a consequence of a design procedure optimizing the dimer structure with
respect to myriad of possible alternative structures rather than to specific
subunit-subunit interactions. This ''design-based stabilization'' is
presently at the speculative level, yet it offers an explanation for the
stability of specific dimers. For instance, the two monomers   of insulin
can form  in  principle two possible adducts  
basically  energetically equivalent in
terms of contact strength\cite{book_proteins}. However, only one adduct 
is found in aqueous solution
and in the crystal phase\cite{whit}.

Furthermore, this work allows also for the first detailed theoretical
characterization of the dynamics of association of a multimeric protein.
First, it is shown that the design of the dimer is very highly optimized at
the interface to stabilize the subunit-subunit interactions, in agreement
with the experimental observation that the quaternary structure of proteins
is disrupted only in drastic conditions or by mutating key residues at the
interface\cite{banci,whit}. Second, we learn that at the first and most
dramatic stages of the dissociation process, only the strongest interfacial
interactions forces are maintained, whereas the intra-monomer interactions
turn out to be almost unaltered. Consistently, we expect that the monomers
do not rearrange significantly upon dissociation, the only region
experiencing large rearrangements being at the interface. This is absolutely
consistent with the recent high resolution structure of the monomeric form
of the dimeric enzyme Cu,Zn superoxide dismutase\cite{banci}, which exhibits
minimal differences with the native protein.

The present results may be of help for protein-engineering experiment and
for a deeper understanding of function/structure relationships of
multimeric proteins.

\section{acknowledgments}
We thank J.R. Banavar for many fruitful discussions and J.N. Onuchic for 
advice and suggestions.
Support from {\it MURST -- Progetti di Ricerca di Rilevante Interesse Nazionale} is
gratefully acknowledged.
During the last three months C.C. has been supported by the 
NSF (Grant \# 96-03839) and by the La Jolla Interfaces in Science 
program (sponsored by the Burroughs Wellcome Fund).

\newpage

\subsection{ Captions to the figures}

{\bf Fig. 1}. Native dimer: (a)\ Structure and numbering (indicated on each
bead) of the type of the bead (as in column 1 of Table\ I); (b)
Inner(squares)- and inter (dots)-chain contacts. In both figures, the chains 
{\it A }and {\it B }are represented in light and dark color, respectively.
In our model, the sequences $A$ and $B$ are chosen not to be necessarily
equal, in order to explore the relevance of the symmetry of the sequence for
protein aggregation.

{\bf Fig. 2. }Energy of the model protein plotted as a function of the
distance {\it d } between subunits.

{\bf Fig. 3.} Subunit dissociation (dark line) and re-association forces
(light line) as resulting from constrained MD simulations where the two
centers of mass of the two subunits are kept at fixed distance {\it r. }The
forces are measured after equilibrium at temperature 0.06 $\varepsilon $ has
been reached.  Multiple points, very close to each other, indicate repeated
calculations and give an estimate of the error bar for the various measured
forces. The ''static'' force, indicated by the black line, is the force
between the monomers immediately after they are separated out at distance 
{\it r} from their native state. The force vs {\it r} curve is then fitted
with the function $F^{fit}(r)=\gamma \{(\frac \rho r)^{\alpha _1}-(\frac \rho
r)^{\alpha _2}\}$ (best fit parameters $\gamma =20.8\epsilon $, $\rho
=2.4\sigma _0$ , $\alpha _1=22.8$ and $\alpha _2=8.5$, correlation
coefficient among fitted and calculated values equal to 0.9998).

{\bf Fig. 4. }Distributions of inter- (light shadow) and intra-monomeric
interactions (dark shadow) normalized to the total number of contacts: at 
{\it r}=15.5 \AA\ (native state), (b) {\it r}=23.4 \AA\ and (c) {\it r}=
30.6 \AA . Notice that stronger interactions are favored at the
subunit-subunit interface with respect of the intramonomer interactions.

{\bf Fig. 5.}{\it \ }Monomer deformation (distance of a monomer from the
corresponding native structure, averaged over the two monomers) as a
function of the distance {\it r }at T= 0.05 $\epsilon $ (a), T=0.15 $%
\epsilon $ (b) and T=0.01 $\epsilon $ (c)$.$

{\bf Fig. 6.} Comparison between the optimized structures of one subunit
folded as in the native protein (dark line) and as independent monomer
(light line). Residues at the interface are indicated with circles. 

{\bf Fig. 7.} Dissociation and re-association forces 
for two dimers with the same native state but different pairs of sequences:
one obtained with an optimal design $(a)$ and the other with a poor design 
quality $(b)$.
Hysteresis loop width correlates well with the degree of dimer stability.
\newpage

%-----------------------------------

%-------------------------------------------------------------
%-------------------------------------------------------------
% ------------- TABLES & TABLE CAPTIONS ------------------
%-------------------------------------------------------------

\begin{table}[t]
\begin{minipage}{0.48\linewidth}
\caption{\scriptsize
Number of amino acids of each kind, occurring in A and B sequences
(column 3), according to their occurrence in proteins (column 2).}
\centering
\begin{tabular}{ccc}
{Residue } & {Occurrence } & {Occurrence }\\
{Type} & { in Proteins (\%)} & {in our model (\#)} \\
\phantom{re}(1) & \phantom{re}(2) & \phantom{re}(3) \\
\hline
 CYS (type 1) &	1.7 &	 1 \\
 MET (type 2) &	2.4 &	 2 \\
 PHE (type 3) &	4.1 &	 3 \\
 ILE (type 4) &	5.8 &	 3 \\
 LEU (type 5) &	9.4 &	 5 \\
 VAL (type 6) &	6.6 &	 4 \\
 TRP (type 7) &	1.2 &	 1 \\
 TYR (type 8) &	3.2 &	 2 \\
 ALA (type 9) &	7.6 &	 4 \\
 GLY (type 10) &	6.8 &	 4 \\
 THR (type 11) &	5.7 &	 3 \\
 SER (type 12) &	7.1 &	 3 \\
 ASN (type 13) &	4.5 &	 2 \\
 GLN (type 14) &	4.0 &	 1 \\
 ASP (type 15) &	5.3 &	 2 \\
 GLU (type 16) &	6.3 &	 3 \\
 HIS (type 17) &	2.2 &	 1 \\
 ARG (type 18) &	5.1 &	 2 \\
 LYS (type 19) &	6.0 &	 2 \\
 PRO (type 20) &	5.0 &	 2 \\
\end{tabular}
%\end{minipage}\hfill
%\begin{minipage}{0.48\linewidth}
%\centering
%\begin{tabular}{|c|c|c|}
%{Residue } & {Occurrence } & {Occurrence }\\
%{Type} & { in Proteins (\%)} & {in our model (\#)} \\
%\phantom{re}(1) & \phantom{re}(2) & \phantom{re}(3) \\
%\hline
%%CYS &	1.7 &	 1 \\
%%MET &	2.4 &	 2 \\
%%PHE &	4.1 &	 3 \\
%%ILE &	5.8 &	 3 \\
%%LEU &	9.4 &	 5 \\
%%VAL &	6.6 &	 4 \\
%%TRP &	1.2 &	 1 \\
%%TYR &	3.2 &	 2 \\
%%ALA &	7.6 &	 4 \\
%%GLY &	6.8 &	 4 \\
%%\hline
%%\end{tabular}

% \end{minipage}
\label{table2}
\end{minipage}
\end{table}

\begin{figure}
\centerline{\psfig{figure=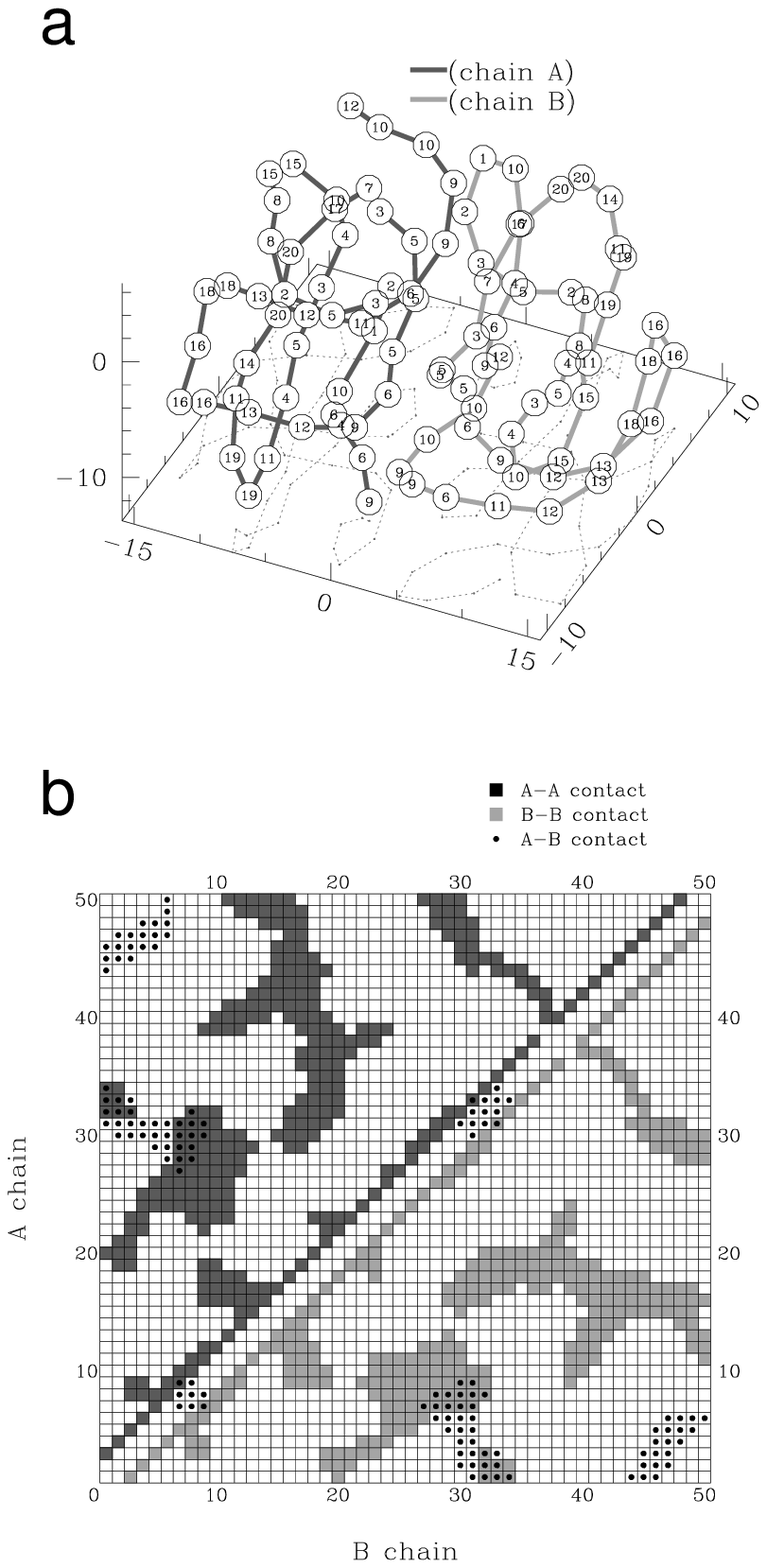,height=20.0cm,angle=0}}
\caption{
}
\end{figure}

\begin{figure}
\centerline{\psfig{figure=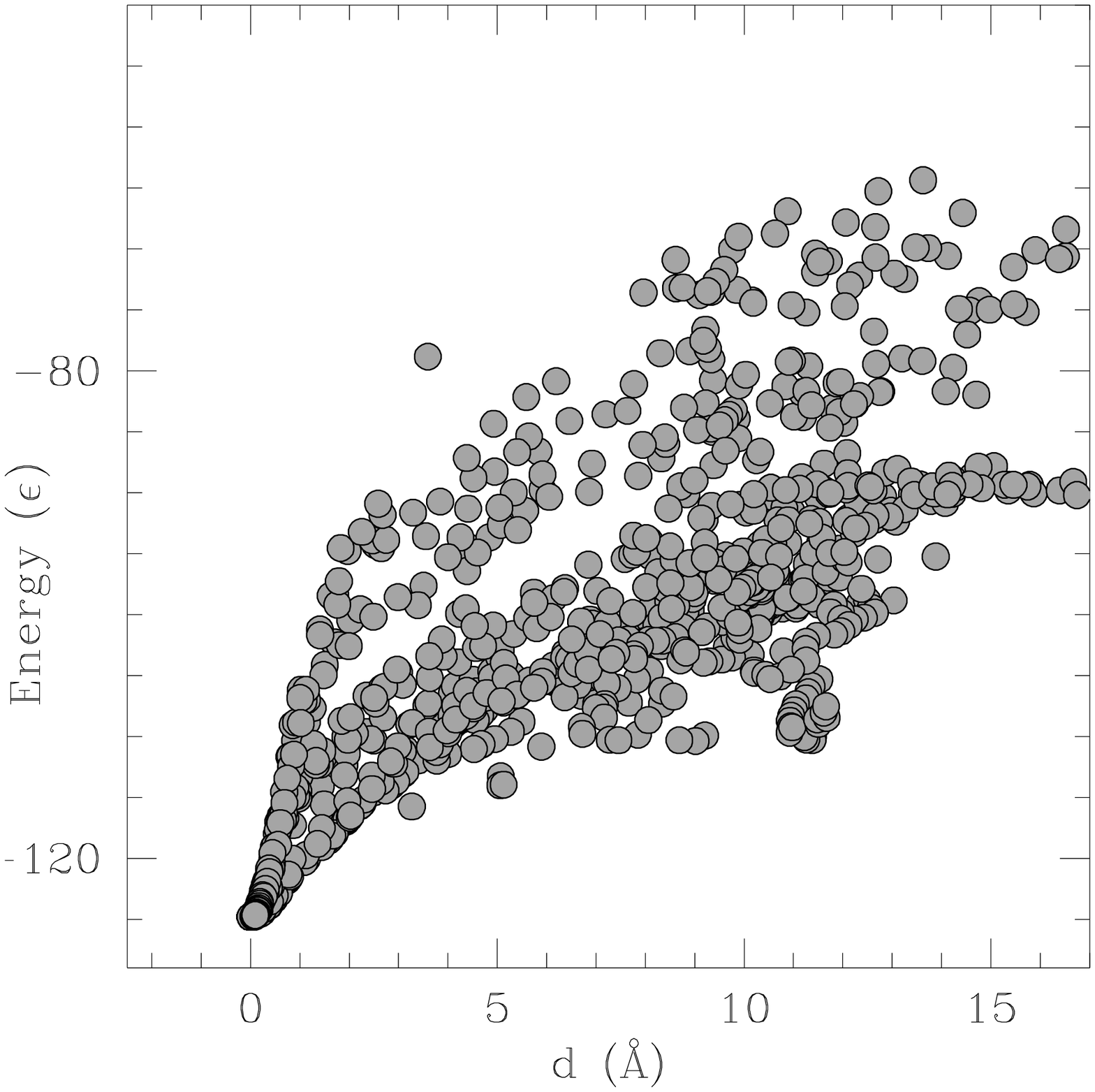,height=10.0cm,angle=0}}
\caption{
}
\end{figure}

\begin{figure}
\centerline{\psfig{figure=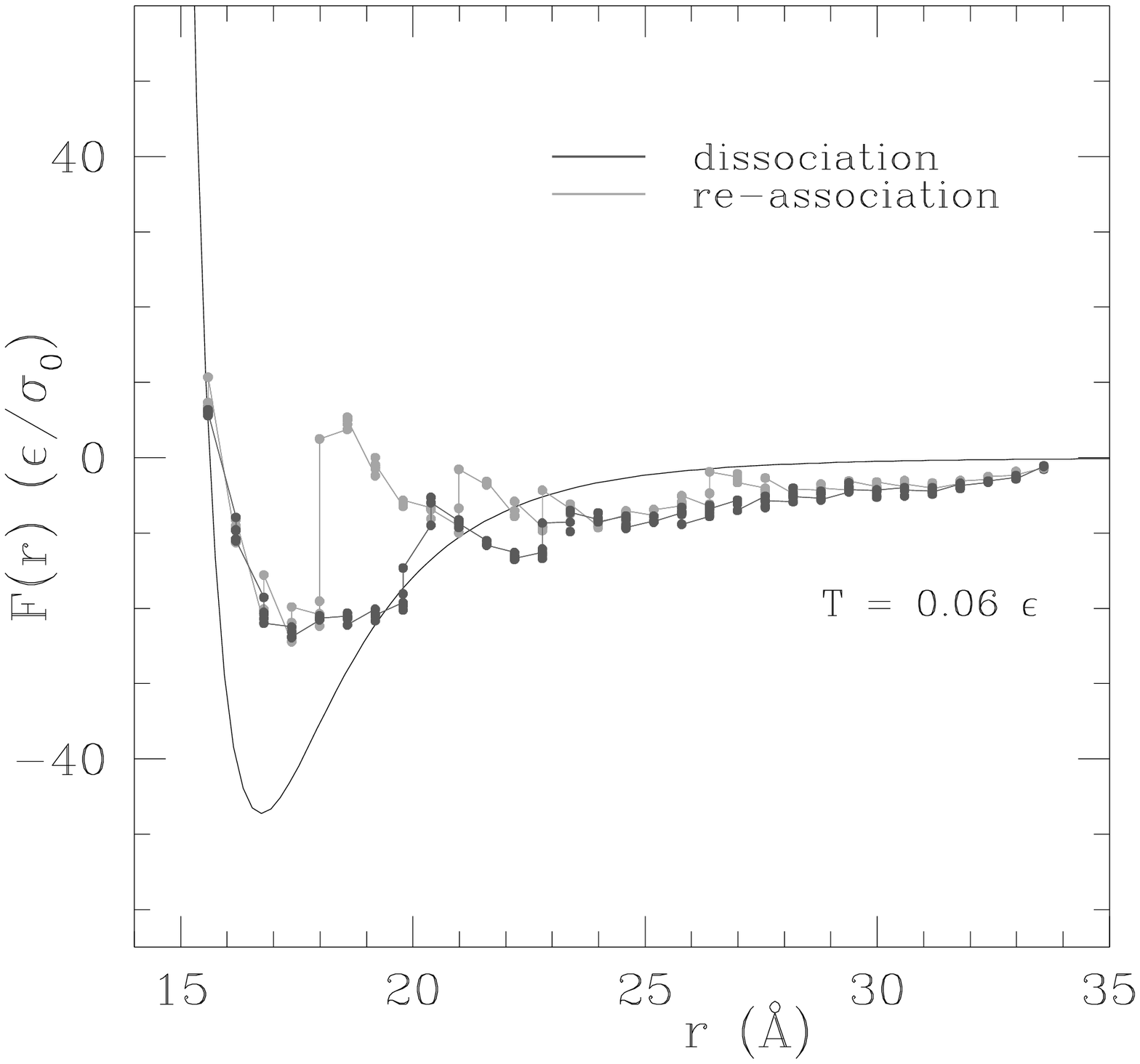,height=10.0cm,angle=0}}
\caption{
}
\end{figure}

\begin{figure}
\centerline{\psfig{figure=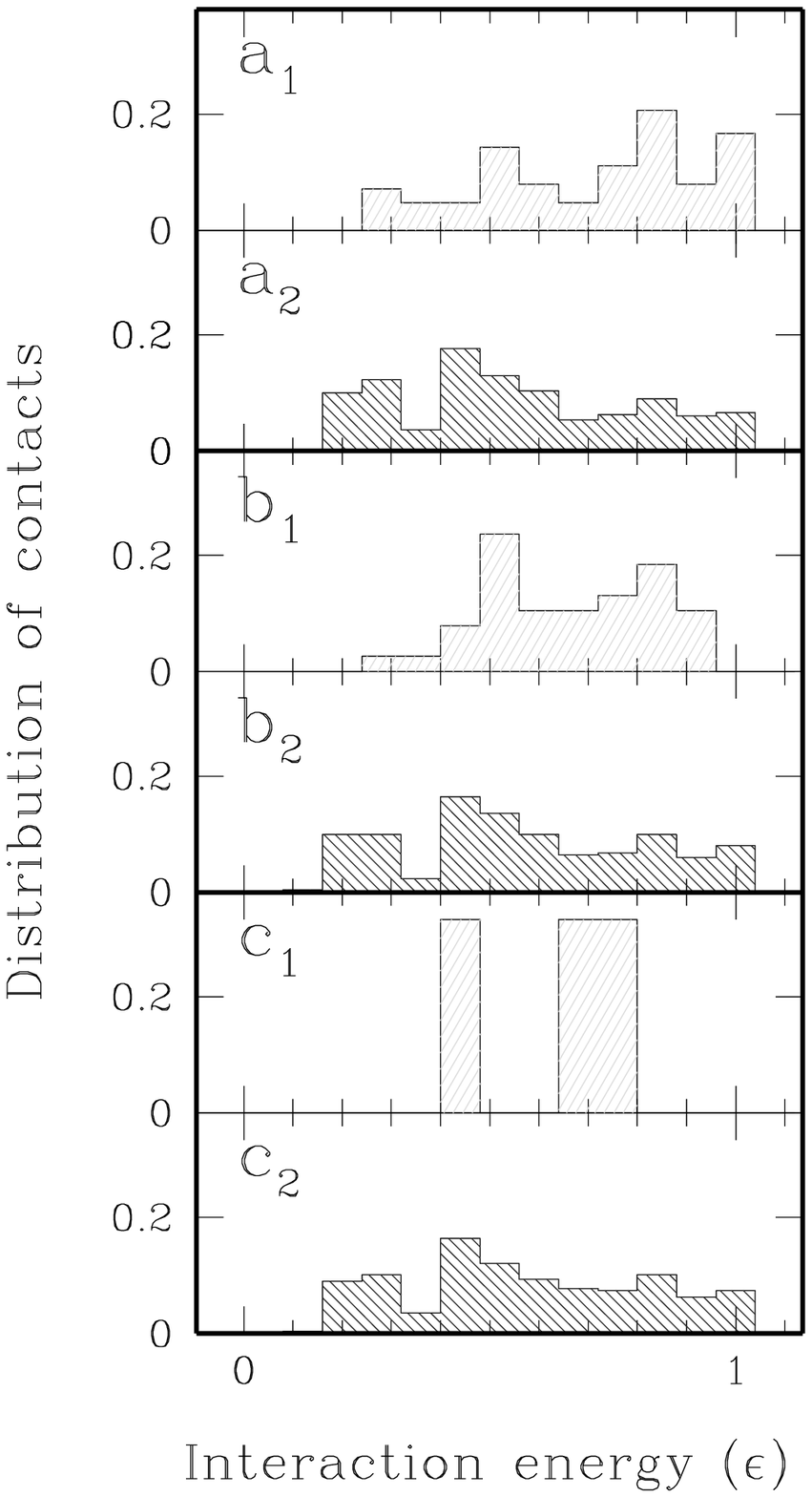,height=15.0cm,angle=0}}
\caption{
}
\end{figure}

\begin{figure}
\centerline{\psfig{figure=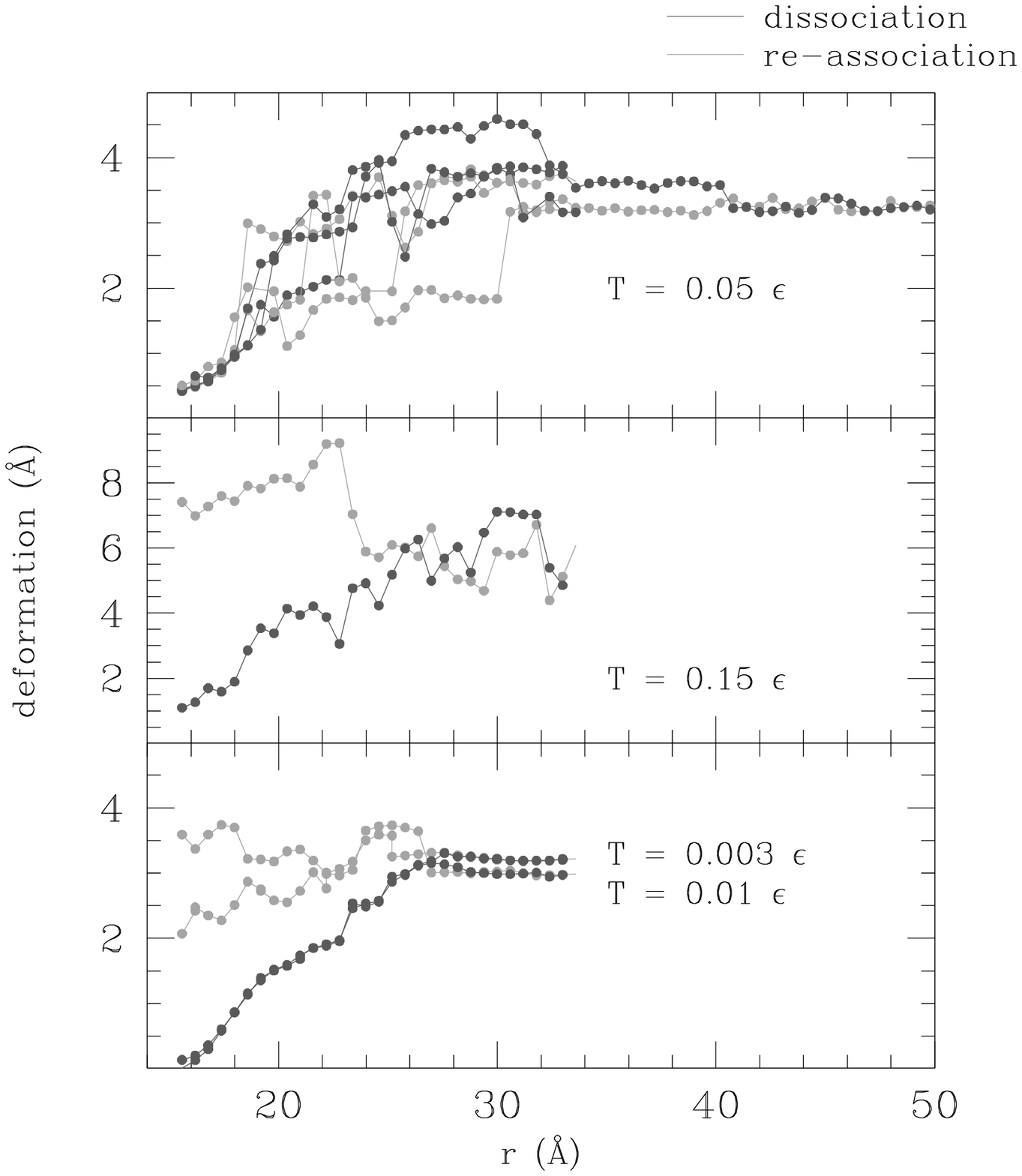,height=10.0cm,angle=0}}
\caption{
}
\end{figure}

\begin{figure}
\centerline{\psfig{figure=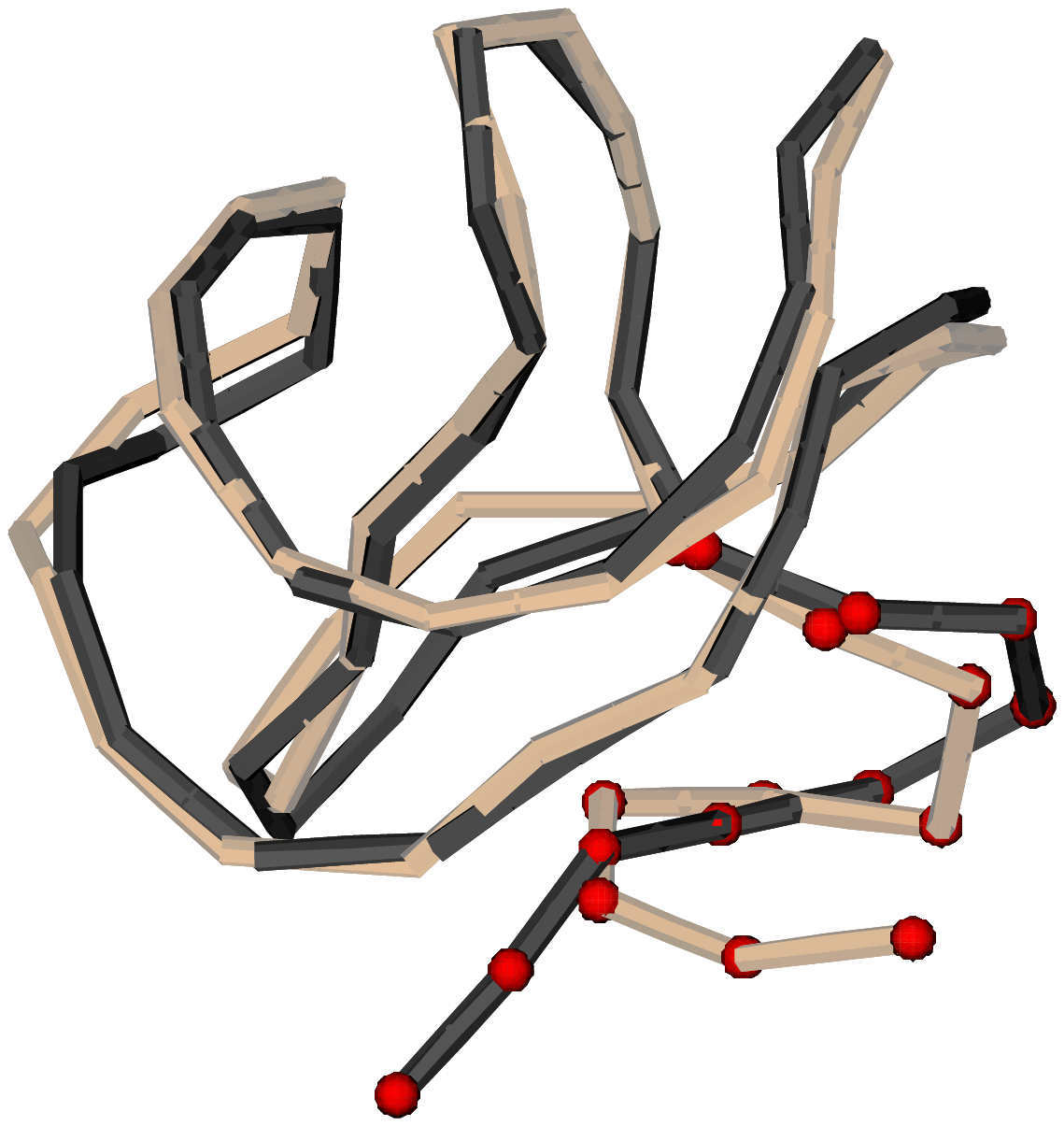,height=17.0cm,angle=0}}
\caption{
}
\end{figure}

\begin{figure}
\centerline{\psfig{figure=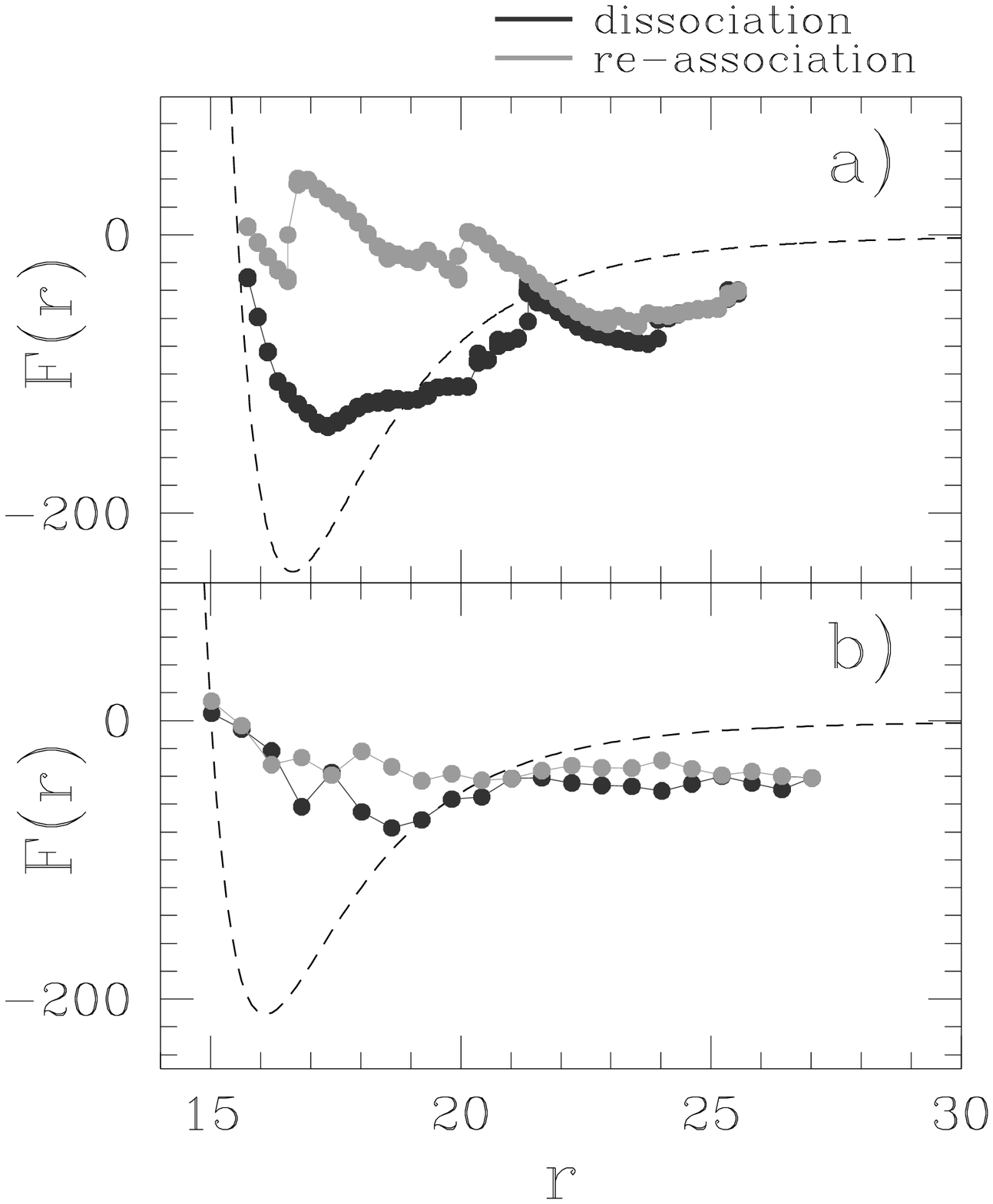,height=10.0cm,angle=0}}
\caption{
}
\end{figure}

\end{document}